\documentclass[10pt]{iopart}

\begin{document}

\title{Eigenvalue correlations in non-Hermitean symplectic random
matrices}

\author{
Eugene Kanzieper \footnote[1]{E-mail address: eugene@phy.cam.ac.uk
/ eugene.kanzieper@weizmann.ac.il} }

\address{Hitachi Cambridge Laboratory, Madingley Road, Cambridge
CB3 0HE, United Kingdom, and \\ Department of Condensed Matter
Physics, Weizmann Institute of Science, Rehovot 76100, Israel}

\begin{abstract}
Correlation function of complex eigenvalues of $N \times N$ random
matrices drawn from non-Hermitean random matrix ensemble of
symplectic symmetry is given in terms of a quaternion determinant.
Spectral properties of Gaussian ensembles are studied in detail in
the regimes of weak and strong non-Hermiticity.
\end{abstract}

\section{Introduction}

Statistical ensembles of generic real, complex and quaternion
matrices have first been introduced in the pioneering work by
Ginibre (1965) who has managed to derive the joint probability
distribution function (j.p.d.f.) of $N$ complex eigenvalues
$\{z_\ell\} = \{x_\ell + i y_\ell\}$ of $N \times N$ complex
($\beta=2$) and quaternion ($\beta=4$) non-Hermitean random
matrices:
\begin{eqnarray}
\label{b=2}
  \hspace{-1cm}
  P_N^{(2)} (z_1,\cdots,z_N) = C_2(N) \prod_{k<\ell} |z_k - z_\ell|^2
                            \prod_{\ell=1}^N w^2(z_\ell,\bar{z}_\ell),
\\
\label{b=4}
    \hspace{-1cm}
    P_N^{(4)} (z_1,\cdots,z_N) = C_4(N)\prod_{k<\ell} |z_k - z_\ell|^2
                            |z_k - \bar{z}_\ell|^2 \prod_{\ell=1}^N
                            |z_\ell - \bar{z}_\ell|^2 \,
w^2(z_\ell,\bar{z}_\ell).
\end{eqnarray}
$C_\beta (N)$ is a normalisation constant, $w^2(z,\bar{z})$ is a
weight function (see discussion below). For real matrices
($\beta=1$) with no further symmetries, the reader is referred to
much later papers by Lehmann and Sommers (1991), and also by
Edelman (1997).

Although Ginibre's derivation of Eqs. (\ref{b=2}) and (\ref{b=4})
holds for random matrices with Gaussian distributed entries, that
is for
\begin{eqnarray}
    \label{gw}
    w^2(z,\bar{z}) = w^2_0(z,{\bar z})= e^{- z {\bar z}},
\end{eqnarray}
we will allow the weight $w^2(z,\bar{z})$ to be an arbitrary
benign function of $z$ and ${\bar z}$ provided the normalisation
$C_\beta(N)$ exists. It should be emphasized \footnote[3]{The
referee is thanked for pointing this out.} that such an innocent
(at first glance) extension is quite nontrivial as it raises the
question about existence of an underlying matrix model whose
eigenvalue representation would coincide with Eqs. (\ref{b=2}) and
(\ref{b=4}). Say, if the underlying non-Hermitean matrix commutes
with its adjoint (such matrices are often called normal matrices,
see, e.g., Oas, 1997), the interpretation of Eqs. (\ref{b=2}) and
(\ref{b=4}) as a non-Gaussian j.p.d.f is correct. Notice, however,
that such a commutativity constraint is not a must though. For
example, an ensemble of weakly non-Hermitean matrices introduced
by Fyodorov, Khoruzhenko and Sommers (1997) is described by
j.p.d.f. of the form (\ref{b=2}) and (\ref{b=4}), see Fyodorov,
Khoruzhenko and Sommers (1998) and Hastings (2000).

Of particular interest is the $n$-point correlation function which
describes a probability density to find $n$ complex eigenvalues
around each of the points $z_1,\cdots,z_n$ while positions of the
remaining levels are unobserved:
\begin{eqnarray}
    \label{n-point}
    \hspace{-1cm}
    R_n^{(\beta)}(z_1,\ldots,z_n) = \frac{N!}{(N-n)!} \int \cdots \int
\,
    d^2 Z_{n+1} \cdots d^2 Z_N \,
    P_N^{(\beta)}(z_1,\cdots,z_N).
\end{eqnarray}
The integration measure $d^2 Z_\ell$ is $d^2 Z_\ell = dx_\ell
dy_\ell$. Quite often, one is also interested in a thermodynamic
limit
\begin{eqnarray}
    \label{magnify}
    \rho_n^{(\beta)}(z_1,\cdots,z_n) = \lim_{N \rightarrow \infty}
    \frac{1}{\delta_N^{2n}}\, R_n^{(\beta)} \left(
    \frac{z_1}{\delta_N},\cdots,\frac{z_n}{\delta_N}
    \right)
\end{eqnarray}
which magnifies spectrum resolution on the appropriate energy
scale $\delta_N$ while letting matrix size $N$ tend to infinity.

At $\beta=2$, the $n$-point correlation function for non-Hermitean
matrix model has also been studied by Ginibre (1965). Adopting the
method of orthogonal polynomials introduced in the context of
Hermitean random matrix theory by Mehta and Gaudin (1960), it is a
straightforward exercise to demonstrate that
$R_n^{(2)}(z_1,\cdots,z_n)$ admits the determinant representation
\begin{eqnarray}
    \label{cf4b=2}
    R_n^{(2)}(z_1,\cdots,z_n) = \det \left[ K_N^{(2)} (z_k,\bar{z}_\ell)
    \right]_{k,\ell=1,\cdots,n}. \nonumber
\end{eqnarray}
The scalar kernel
\begin{eqnarray}
    K_N^{(2)}(z,z^\prime) = w(z,\bar{z}) \, w(z^\prime, \bar{z}^\prime)
    \sum_{k=0}^{N-1} P_k(z) P_k(z^\prime) \nonumber
\end{eqnarray}
is expressed in terms of polynomials $P_k(z)$ orthonormal in the
complex plane $z=x+iy$
\begin{eqnarray}
    \int d^2 Z \, w^2(z,\bar{z}) \, P_k(z) P_\ell(\bar{z}) =
\delta_{k\ell}
    \nonumber
\end{eqnarray}
with respect to the measure $w^2(z,\bar{z}) \, d^2 Z$.

For instance, the density of states and the two-point correlation
function equal
\begin{eqnarray}
    \label{r1=2}
    R_1^{(2)}(z) = K_N(z,\bar{z}) \nonumber
\end{eqnarray}
and
\begin{eqnarray}
    \label{r2=2}
    R_2^{(2)}(z_1,z_2) = K_N(z_1,\bar{z}_1) K_N(z_2,\bar{z}_2)
    - \left| K_N(z_1,\bar{z}_2) \right|^2, \nonumber
\end{eqnarray}
respectively.

Non-Hermitean random matrices at $\beta=4$ have also received some
attention in both physical and mathematical literature especially
following a recent burst of interest to spectral properties of
non-Hermitean random operators (see, e.g., Efetov (1997)).

M. L. Mehta (1967) considered a non-Hermitean matrix model of
symplectic symmetry for Ginibre's weight function
$w^{2}_0(z,\bar{z})=\exp(-z\bar{z})$, and established a quaternion
determinant structure of one- and two-point correlation functions.
He also conjectured a similar structure to hold for all $n$-point
correlation functions. In Ginibre's case, these appear in the
revised 1991 edition of the Mehta's book.

Further progress has come with development of field theoretic
techniques. Kolesnikov and Efetov (1997), driven by possible
applications in quantum chromodynamics (Halasz, Osborn and
Verbaarschot, 1997), have formulated a nonlinear supersymmetry
$\sigma$-model for this class of random matrices, and derived an
expression for the eigenvalue density in a somewhat richer model
[see Eq. (\ref{weight}) below].

More recently, yet another field theory approach (aka replica
method) was outlined by Nishigaki and Kamenev (2002). There, the
well known Mehta's expressions were reproduced for one-point
correlation function in case of the very same Ginibre's weight
function $w^{2}_0(z,\bar{z})$. Unfortunately, both mentioned
techniques run into obstacles when one attempts to study
higher-order correlation functions whilst replica $\sigma$-models
seem to reliably provide asymptotic expansions only.

A different route has been chosen by Hastings (2000) who suggested
that there exists a mapping of non-Hermitean random matrices of
symplectic symmetry onto a fermion field theory. Even though the
method might have been potentially applicable to a study of
$n$-point correlation functions in the {\it bulk} of a complex
spectrum, these have not explicitly been worked out beyond the
two-point correlation function.

Our paper reports on a comprehensive treatment of integrable
structure of non-Hermitean random matrix models at $\beta=4$. It
sets a transparent and coherent framework to study all $n$-point
correlation functions: while easily applied to reproduce the
results of aforementioned studies and extend them to higher-order
correlation functions in the spectrum bulk, it may go much farther
and serve as a proper starting point to explore eigenvalue
correlations near the spectrum edges and/or address the issue of
universality (for a recent review of the universality phenomenon
in the context of Hermitean random matrix models see, e.g.,
Kanzieper and Freilikher, 1999).

The paper is organised as follows. Section 2 announces a most
general form of $n$-point correlation function whatever the weight
function in Eq. (\ref{b=4}) is. A proof is given in Section 3. A
concept of skew orthogonal polynomials which are central to
performing explicit calculations is elaborated in detail in
Section 4. There, exact expressions for skew orthogonal
polynomials are given in terms of multi-fold integrals. For the
Gaussian weight Eq. (\ref{weight}), the polynomials are evaluated
explicitly. In Section 5, $n$-point correlation functions for the
Gaussian weight [Eq. (\ref{weight})] are derived for finite $N$ as
well as in the large-$N$ limit. Section 6 contains concluding
remarks and briefly mentions further possible applications of the
formalism developed.

\section{Correlation function at $\beta=4$ and eigenvalue depletion
along real axis}

For symplectic ensemble, the following representation holds for
$n$-point correlation function:
\begin{eqnarray}
\label{cf4b=4}
    R_n^{(4)}(z_1,\ldots,z_n) = Q\det \left[
    K_N^{(4)}(z_k,z_\ell)
    \right]_{k,\ell=1,\cdots,n}.
\end{eqnarray}
Here $Q\det$ stands for a quaternion determinant (Dyson, 1972).
The $2 \times 2$ matrix kernel
\begin{eqnarray}
    \label{kernel4b=4}
    K_N^{(4)}(z,z^\prime) &=& (\bar{z} - z)^{1/2}
    (\bar{z}^\prime - z^\prime)^{1/2} w(z,\bar{z})\,
    w(z^\prime,\bar{z}^\prime)
    \nonumber \\
    &\times& \left(
    \matrix{
    \kappa_N(\bar{z},z^\prime) &
    -\kappa_N(\bar{z},\bar{z}^\prime) \cr
    \kappa_N(z,z^\prime) &
    -\kappa_N(z,\bar{z}^\prime)
    }
    \right),
\end{eqnarray}
where the `prekernel' $\kappa_N$ is
\begin{eqnarray}
    \label{prekernel}
    \kappa_N(z,z^\prime) = \sum_{k,\ell=0}^{2N-1} p_k(z)\,
    \left(M^{-1}\right)_{k\ell}
    \, p_\ell(z^\prime),
\end{eqnarray}
and $M^{-1}$ is an inverse to the real antisymmetric matrix $M$
with the entries
\begin{eqnarray}
    \label{M-matrix}
    M_{k\ell} = \int d^2 Z \, (\bar{z}-z)\, w^2(z,\bar{z})\,
    \left[
    p_k(z) p_\ell(\bar{z}) - p_\ell(z) p_k(\bar{z})
    \right].
\end{eqnarray}

The polynomials $p_k(z)$ are arbitrary provided the inverse
$M^{-1}$ exists. Since the matrix $M$ is antisymmetric, the
formulas would be simplest had it contained $N$ copies of the $2
\times 2$ matrix
\begin{eqnarray}
    i\sigma_y = \left(
    \matrix{
    0  & 1 \cr
    -1 & 0
    }
    \right) \nonumber
\end{eqnarray}
along the main diagonal. This is achieved by letting $p_k(z)$ be
skew-orthogonal polynomials $q_k(z)$ in the complex domain:
\begin{eqnarray}
    \label{skew-orthogonality-a}
    \langle q_{2k+1},q_{2\ell} \rangle_S &=& - \langle
q_{2\ell},q_{2k+1}
    \rangle_S = r_k \delta_{k\ell}, \\
    \label{skew-orthogonality-b}
    \langle q_{2k+1},q_{2\ell+1} \rangle_S &=& \langle
    q_{2k},q_{2\ell}
    \rangle_S = 0.
\end{eqnarray}
The skew product $\langle f, g \rangle_S$ is defined as
\footnote[6]{Notice a difference from the skew orthogonality
arising in the context of Hermitean random matrices (Mahoux and
Mehta, 1991).}
\begin{eqnarray}
    \label{skew-product}
    \langle f,g \rangle_S = \int d^2 Z \, (\bar{z}-z)\, w^2(z,\bar{z})\,
    \left[
    f(z) g(\bar{z}) - f(\bar{z}) g(z)
    \right].
\end{eqnarray}
With this choice in mind, the prekernel $\kappa_N$ further
simplifies to
\begin{eqnarray}
    \label{prekernel-skew}
    \kappa_N(z,z^\prime) = \sum_{k=0}^{N-1}
    \frac{q_{2k+1}(z)\, q_{2k}(z^\prime) - q_{2k+1}(z^\prime)\,
    q_{2k}(z)}{r_k}.
\end{eqnarray}

In particular, the density of states and the two-point correlation
functions are expressed as
\begin{eqnarray}
    \label{r1=4}
    R_1^{(4)}(z) = (\bar{z}-z)\, w^2(z,\bar{z})\,
    \kappa_N(z,\bar{z})
\end{eqnarray}
and
\begin{eqnarray}
    \label{r2=4}
    \hspace{-1cm}
    R_2^{(4)}(z_1,z_2) &=& (\bar{z}_1 - z_1)(\bar{z}_2 - z_2)\,
    w^2(z_1,\bar{z}_1)\,w^2(z_2,\bar{z}_2) \nonumber \\
    &\times& \left[
    \kappa_N(z_1,\bar{z}_1) \kappa_N(z_2,\bar{z}_2)
     -
    \left| \kappa_N(z_1,z_2)
    \right|^2
    +
    \left| \kappa_N(z_1,\bar{z}_2)
    \right|^2
    \right],
\end{eqnarray}
respectively.

Notice that, in accordance with our solution [Eqs. (\ref{cf4b=4}) and
(\ref{kernel4b=4})], the $n$-point correlation function
universally vanishes along the real axes $\Im {\rm m}\, z_\ell
=0$,
\begin{eqnarray}
    R_{n}^{(4)}(z_{1},\cdots,z_{n}) \propto \prod_{\ell=1}^{n}\,
    \left[\Im {\rm m}\,z_{\ell} \right]^\alpha, \;\;\;
    \alpha\equiv 2 \nonumber
\end{eqnarray}
whatever the weight function $w^{2}(z,\bar{z})$ is. It is this
specific feature of spectral correlations in symplectic ensembles
of non-Hermitean random matrices that has been revealed, for
$n=1$, in early numerical simulations due to Halasz, Osborn and
Verbaarschot (1997). Qualitatively, such a depletion of complex
eigenvalues along the real axis might have been anticipated after
a brief inspection of both the j.p.d.f. [Eq. (\ref{b=4})] and the
definition of $n$-point correlation function [Eq.
(\ref{n-point})].

The results announced [Eqs. (\ref{cf4b=4}) --
(\ref{prekernel-skew})] will be derived in Section 3. In Section
4, we study properties of skew-orthogonal polynomials which
constitute a natural basis to perform actual calculations of
spectral fluctuations in $\beta=4$ non-Hermitean random matrix
ensembles. The latter are addressed in Section 5, where we
consider an ensemble of $N \times N$ random matrices associated with a
Gaussian measure. Correlation functions in the regimes of strong
(Ginibre, 1965) and weak
(Fyodorov, Khoruzhenko and Sommers, 1997) non-Hermiticity are
explicitly given there for finite $N$ as
well as in the limit of infinite matrices.

\section{Derivation}

To derive a quaternion determinant representation of $n$-point
correlation function, we will follow an elegant idea of Tracy and
Widom (1998). These authors have introduced generating functional
\begin{eqnarray}
    G[f] = \int \cdots \int \,
    d^2 Z_1 \cdots d^2 Z_N \,
    P_N (z_1, \cdots, z_N)
    \,
    \prod_{k=1}^N [1 + f (z_k)] \nonumber
\end{eqnarray}
such that the $n$-point correlation function $R_n(z_1,\cdots,z_n)$
defined by Eq. (\ref{n-point}) can be viewed as the coefficient of
$\alpha_1 \cdots \alpha_n$ in the expansion of $G[f]$ for a
particular choice $f(z) = \sum_{r=1}^N \alpha_r \,
\delta^2(z-z_r)$. Assuming that $G[f]$ admits the representation
\begin{eqnarray}
G[f] = \sqrt{{\rm det}(I + K_N f)}, \nonumber
\end{eqnarray}
where $K_N$ denotes the operator with $2\times 2$ (self-dual)
matrix kernel $K_N(z,z^\prime)$ and $f$ denotes multiplication by
that function, Tracy and Widom have explicitly evaluated the
coefficient of $\alpha_1 \cdots \alpha_n$ and found it to be equal
to the quaternion determinant in the r.h.s. of Eq. (\ref{cf4b=4}).

Hence, in accordance with this statement (which we will name the
Tracy-Widom theorem) one has to seek a suitable representation for
$G[f]$ with the j.p.d.f. given by Eq. (\ref{b=4}). This is easy.
Due to the identity
\begin{eqnarray}
    %\hspace{-2cm}
    \prod_{k<\ell}(x_k - x_\ell) \,
    \prod_{k,\ell=1}^N (y_k - x_\ell) \,
    \prod_{k>\ell}(y_k - y_\ell)
    =
    \det \left[
    \matrix{
    x_\ell^{k-1} \cr
    y_\ell^{k-1}}
    \right]_{
    \matrix
    {
    \scriptstyle k=1,\cdots,2N\hfill \cr
    \scriptstyle \ell=1,\cdots,N\hfill\cr
    }
    } \nonumber
\end{eqnarray}
one notices that
\begin{eqnarray}
    %\hspace{-2cm}
    \prod_{k<\ell} |z_k - z_\ell|^2
    |z_k - \bar{z}_\ell|^2
    \prod_{\ell=1}^N
    (\bar{z}_\ell - z_\ell)
    =
    \det \left[
    \matrix{
    z_\ell^{k-1} \cr
    \bar{z}_\ell^{k-1}}
    \right]_{
    \matrix
    {
    \scriptstyle k=1,\cdots,2N\hfill \cr
    \scriptstyle \ell=1,\cdots,N\hfill\cr
    }
    }. \nonumber
\end{eqnarray}
To derive the latter, we have put $x_{\ell} = z_{\ell}$
and $y_{\ell} = \bar{z}_{\ell}$ in the former. With this result in
mind, the j.p.d.f. $P_N^{(4)}$ can be cast into the form
\begin{eqnarray}
\label{pn4-det}
    P_N^{(4)} (z_1,\cdots,z_N) &=& C_4(N) \prod_{\ell=1}^N
                            (z_\ell - \bar{z}_\ell) \,
w^2(z_\ell,\bar{z}_\ell)
    \nonumber \\
    &\times& \det \left[
    \matrix{
    p_{k-1}(z_\ell) \cr
    p_{k-1}(\bar{z}_\ell)}
    \right]_{
    \matrix
    {
    \scriptstyle k=1,\cdots,2N\hfill \cr
    \scriptstyle \ell=1,\cdots,N\hfill\cr
    }
    }. \nonumber
\end{eqnarray}
Here, we have replaced the sequence of monomials
$\{z_{\ell}^{k}\}$ by arbitrary monic polynomials
$\{p_k(z_\ell)\}$ of degree $k$ as this leaves the value of
determinant intact. If the polynomials $p_k$ were not monic, the
normalisation prefactor $C_4(N)$ would change.

This representation is fairly useful due to de Bruijn's (1955)
integration formula
\begin{eqnarray}
    \label{dB}
    \int \cdots \int d\alpha (Z_1) \cdots d\alpha (Z_N)
    \, \det
    \left[
    \matrix{
    f_{k}(z_\ell) \cr
    g_{k}(z_\ell)}
    \right]_{
    \matrix
    {
    \scriptstyle k=1,\cdots,2N\hfill \cr
    \scriptstyle \ell=1,\cdots,N\hfill\cr
    }
    } \nonumber \\
    = (2N)! \, {\rm Pf}
    \left[
    \int d \alpha (Z) \,
    [f_{k}(z) g_{\ell}(z) - f_\ell(z) g_k(z)]
    \right]_{k,\ell=1,\cdots,2N} \nonumber
\end{eqnarray}
in which $d\alpha(Z)$ is an integration measure
and `Pf' stands for pfaffian. One derives the following
chain:
\begin{eqnarray}
    \label{g2f-a}
    G^2[f]
    &\propto&
    \det
    \bigg[
    \int d^2 Z \, (\bar{z}-z) \, w^2(z,\bar{z})
    \nonumber \\
    &\times&[
    1+f(z)]
    \,
    [p_k(z)\, p_\ell(\bar{z}) - p_\ell(z)\, p_k(\bar{z})]
    \bigg] \nonumber
    \\
    \label{g2f-b}
    &\propto&
    \det
    \bigg[
    M_{k\ell} + \int d^2 Z \, (\bar{z}-z) \, w^2(z,\bar{z}) \nonumber \\
    &\times&
    f(z)
    \,
    [p_k(z)\, p_\ell(\bar{z}) - p_\ell(z)\, p_k(\bar{z})]
    \bigg]
    \\
    \label{g2f-c}
    &\propto&
    \det
    \bigg[
    \delta_{k\ell} + \int d^2 Z \, (\bar{z}-z)
    w^2(z,\bar{z}) \,
    \nonumber \\
    &\times&
    f(z)
    \,
    [\pi_k(z)\, p_\ell(\bar{z}) - p_\ell(z)\, \pi_k(\bar{z})]
    \bigg].
\end{eqnarray}
Matrix $M$ arisen in Eq. (\ref{g2f-b}) is defined by Eq.
(\ref{M-matrix}). To derive Eq. (\ref{g2f-c}), we have factored
out $M$ on the left. Notice that Eq. (\ref{g2f-c}) automatically
bears a proper normalisation $G^{2}[0] \equiv 1$. The polynomials
$\pi_k$ are
\begin{eqnarray}
    \pi_k(z) = \sum_{\ell=0}^{2N-1} \, (M^{-1})_{k\ell}\, p_\ell(z).
\nonumber
\end{eqnarray}

The matrix appearing under the sign of determinant in Eq.
(\ref{g2f-c}) can be represented as $I+AB$ with
\begin{eqnarray}
    A(\ell,z) &=& f(z) (\bar{z} -z)^{1/2} w(z,\bar{z})
    \left(
    \pi_{\ell}(z), - \pi_{\ell}(\bar{z})
    \right), \nonumber \\
    B(z,\ell) &=& (\bar{z}-z)^{1/2} w(z,\bar{z})
    \left(
    \matrix{
    p_{\ell}(\bar{z})  \cr
    p_{\ell}(z)
    }
    \right). \nonumber
\end{eqnarray}
As soon as the transposition does not affect the value of the
determinant, one observes the identity $\det(I+AB) = \det(I+BA)$.
In this case $BA$ is the integral operator with matrix kernel
$K_N(z_1,z_2)f(z_2)$ where
\begin{eqnarray}
    K_N(z_1,z_2) &=&
    (\bar{z}_1 - z_1)^{1/2} \,(\bar{z}_2 - z_2)^{1/2}
    w(z_1,\bar{z}_1) w(z_2,\bar{z}_{2}) \nonumber \\
    &\times&
    \left(
    \matrix{
    \sum_\ell p_{\ell}(\bar{z}_1) \pi_{\ell}(z_2) &
    -\sum_\ell p_{\ell}(\bar{z}_1) \pi_{\ell}(\bar{z}_2)\cr
    \sum_\ell p_{\ell}(z_1) \pi_{\ell}(z_2) &
    -\sum_\ell p_{\ell}(z_1) \pi_{\ell}(\bar{z}_2)
    }
    \right). \nonumber
\end{eqnarray}
Hence, we have proven that $G^2[f] = \det(I + K_N f)$. Since the
$2\times 2$ matrix $K_N$ is self-dual\footnote[4] {Indeed, since
the quaternion $\kappa_N$ is represented by $2\times 2$ matrix
\begin{eqnarray}
    \theta[\kappa_{N}] =
    \left(
    \matrix{
    \kappa_{N}(\bar{z}_1,z_{2}) &
    -\kappa_{N}(\bar{z}_{1},\bar{z}_{2}) \cr
    \kappa_{N}(z_{1},z_{2}) &
    -\kappa_{N}(z_1,\bar{z}_{2})
    }
    \right), \nonumber
\end{eqnarray}
the dual quaternion $\tilde{\kappa}_{N}$ is given by
\begin{eqnarray}
    \theta[\tilde{\kappa}_{N}] =
    \left(
    \matrix{
    -\kappa_{N}(z_1,\bar{z}_{2}) &
    \kappa_{N}(\bar{z}_{1},\bar{z}_{2}) \cr
    -\kappa_{N}(z_{1},z_{2}) &
    \kappa_{N}(\bar{z}_1,z_{2})
    }
    \right). \nonumber
\end{eqnarray}
Self-duality is a consequence of the equality $\sigma_y
\theta[\tilde{\kappa}_{N}^{T}] = \theta[\kappa_{N}]\sigma_{y}$. },
Eq. (\ref{cf4b=4}) follows by virtue of Tracy-Widom theorem. This
completes our proof.

\section{Skew-orthogonal polynomials}

\subsection{{\bf General weight} $w^2(z,\bar{z})$}

We have seen in Section 2 that skew-orthogonal polynomials defined
by Eqs. (\ref{skew-orthogonality-a}) -- (\ref{skew-product})
represent a natural basis in which calculations become simplest.
These can explicitly be found for a general weight function
$w^2(z,\bar{z})$ provided the integrals below make sense and, in a
monic normalisation, are given by the following $2n$-fold
integrals:
\begin{eqnarray}
    \label{even-matrix}
    \hspace{-1cm}
    q_{2n}(z) &\equiv& \frac{1}{A_n}
    \int \cdots \int \,
    d^{2} Z_{1} \cdots d^{2} Z_{n}
    \prod_{\ell=1}^{n}\, (z-z_{\ell}) (z - \bar{z}_{\ell}) \nonumber \\
    &\times& \prod_{k<\ell} |z_k - z_\ell|^2
    |z_k - \bar{z}_\ell|^2 \prod_{\ell=1}^n
    |z_\ell - \bar{z}_\ell|^2 \, w^2(z_\ell,\bar{z}_\ell), \\
    \label{odd-matrix}
    \hspace{-1cm}
    q_{2n+1}(z) &\equiv& \frac{1}{A_n}
    \int \cdots \int \,
    d^{2} Z_{1} \cdots d^{2} Z_{n}
    \prod_{\ell=1}^{n}\, (z-z_{\ell}) (z - \bar{z}_{\ell}) \nonumber \\
    &\times&
    \left(
    z + \sum_{k=1}^{n} (z_{k}+\bar{z}_{k}) + c_{n}
    \right) \, \nonumber \\
    &\times&
    \prod_{k<\ell} |z_k - z_\ell|^2
    |z_k - \bar{z}_\ell|^2 \prod_{\ell=1}^n
    |z_\ell - \bar{z}_\ell|^2 \, w^2(z_\ell,\bar{z}_\ell).
\end{eqnarray}
Here $c_{n}$ is an arbitrary constant which we will set to zero,
$c_n=0$ whilst
\begin{eqnarray}
    \label{An}
    \hspace{-2.2cm}
    A_n =
    \int \cdots \int \,
    d^{2} Z_{1} \cdots d^{2} Z_{n}
    \, \prod_{k<\ell} |z_k - z_\ell|^2
    |z_k - \bar{z}_\ell|^2 \prod_{\ell=1}^n
    |z_\ell - \bar{z}_\ell|^2 \, w^2(z_\ell,\bar{z}_\ell).
\end{eqnarray}
For similar representations of skew-orthogonal
polynomials arisen in the context of Hermitean random matrix
theory see Eynard (2001).

To prove that Eqs. (\ref{even-matrix}) and (\ref{odd-matrix}) obey
skew-orthogonality relations [Eqs. (\ref{skew-orthogonality-a}) --
(\ref{skew-product})], it is sufficient to show that (i) $\langle
q_{2n},\, z^{m}\rangle_{S} = 0$ and (ii) $\langle q_{2n+1}, \,
z^{m}\rangle_{S} = 0$ for integer $0 \le m \le 2n-1$.

(i) Consider
\begin{eqnarray}
    \hspace{-1cm}
    \langle q_{2n}, \, z^{m} \rangle_{S}
    &\propto&
    \int d^{2}Z \, (\bar{z}-z) \, w^{2}(z,\bar{z}) \,
    \int \cdots \int \,
    d^{2}Z_{1} \cdots d^{2} Z_{n} \nonumber \\
    &\times&
    \prod_{k<\ell} |z_k - z_\ell|^2
    |z_k - \bar{z}_\ell|^2 \prod_{\ell=1}^n
    |z_\ell - \bar{z}_\ell|^2 \, w^2(z_\ell,\bar{z}_\ell) \nonumber \\
    &\times&
    \left[
    {\bar{z}}^{m} \, \prod_{\ell=1}^{n}\, (z-z_{\ell}) (z -
\bar{z}_{\ell})
    -
    z^{m}
    \, \prod_{\ell=1}^{n}\, (\bar{z}-z_{\ell}) (\bar{z} -
\bar{z}_{\ell})
    \right]. \nonumber
\end{eqnarray}
As soon as
\begin{eqnarray}
    \prod_{k<\ell} |z_k - z_\ell|^2
    |z_k - \bar{z}_\ell|^2 \prod_{\ell=1}^n
    (\bar{z}_\ell - z_\ell)
    \,
    \prod_{\ell=1}^n \,
    (z-z_{\ell}) (z - \bar{z}_{\ell}) \nonumber \\
    =
    \det \left(
    \matrix{
    1 & z_{1} & \cdots & z_{1}^{2n-1} & z_{1}^{2n} \cr
    1 & \bar{z}_{1} & \cdots & \bar{z}_{1}^{2n-1}& \bar{z}_{1}^{2n} \cr
    \vdots & \vdots & & \vdots & \vdots \cr
    1 & z_{n} & \cdots & z_{n}^{2n-1} & z_{n}^{2n} \cr
    1 & \bar{z}_{n} & \cdots & \bar{z}_{n}^{2n-1}& \bar{z}_{n}^{2n} \cr
    1 & z & \cdots & z^{2n-1}& z^{2n}
    }
    \right), \nonumber
\end{eqnarray}
we derive
\begin{eqnarray}
    \hspace{-1cm}
    \langle q_{2n}, \, z^{m} \rangle_{S}
    &\propto&
    \int \cdots \int \,
    d^{2}Z_{1} \cdots d^{2}Z_{n} \, d^{2}Z_{n+1} \,
    \prod_{\ell=1}^{n+1}
    \,
    (z_{\ell} - \bar{z}_{\ell})\, w^{2}(z_{\ell},\bar{z}_{\ell})
    \nonumber \\
    &\times& \det
    \left(
    \matrix{
    1 & z_{1} & \cdots & z_{1}^{2n} & 0 \cr
    1 & \bar{z}_{1} & \cdots & \bar{z}_{1}^{2n}& 0 \cr
    \vdots & \vdots & & \vdots & \vdots \cr
    1 & z_{n} & \cdots & z_{n}^{2n} & 0 \cr
    1 & \bar{z}_{n} & \cdots & \bar{z}_{n}^{2n}& 0 \cr
    1 & z_{n+1} & \cdots & z_{n+1}^{2n}& z_{n+1}^{m} \cr
    1 & \bar{z}_{n+1} & \cdots & \bar{z}_{n+1}^{2n}& \bar{z}_{n+1}^{m}
    }
    \right), \nonumber
\end{eqnarray}
where we have introduced $z_{n+1} = z$. Since a particular
enumeration of $z_\ell$ ($1 \le \ell \le n+1$) is irrelevant, this
reduces to
\begin{eqnarray}
    \hspace{-1cm}
    \langle q_{2n}, \, z^{m} \rangle_{S}
    &\propto&
    \int \cdots \int \,
    d^{2}Z_{1} \cdots d^{2}Z_{n} \, d^{2}Z_{n+1} \,
    \prod_{\ell=1}^{n+1}
    \,
    (z_{\ell} - \bar{z}_{\ell})\, w^{2}(z_{\ell},\bar{z}_{\ell})
    \nonumber \\
    &\times& \det
    \left(
    \matrix{
    1 & z_{1} & \cdots & z_{1}^{2n} & z_{1}^{m} \cr
    1 & \bar{z}_{1} & \cdots & \bar{z}_{1}^{2n}& \bar{z}_{1}^{m} \cr
    \vdots & \vdots & & \vdots & \vdots \cr
    1 & z_{n+1} & \cdots & z_{n+1}^{2n}& z_{n+1}^{m} \cr
    1 & \bar{z}_{n+1} & \cdots & \bar{z}_{n+1}^{2n}& \bar{z}_{n+1}^{m}
    }
    \right). \nonumber
\end{eqnarray}
The latter integrand obviously vanishes for $0 \le m \le 2n$ thus
completing the proof of Eq. (\ref{even-matrix}).

(ii) Consider
\begin{eqnarray}
    \hspace{-1cm}
    \langle q_{2n+1},\, z^{m} \rangle_{S}
    &\propto&
    \int d^{2}Z \, (\bar{z}-z) \, w^{2}(z,\bar{z}) \,
    \int \cdots \int \,
    d^{2}Z_{1} \cdots d^{2} Z_{n} \nonumber \\
    &\times&
    \prod_{k<\ell} |z_k - z_\ell|^2
    |z_k - \bar{z}_\ell|^2 \prod_{\ell=1}^n
    |z_\ell - \bar{z}_\ell|^2 \, w^2(z_\ell,\bar{z}_\ell) \nonumber \\
    &\times&
    \left[
    {\bar{z}}^{m} \,
    \left(
    z + \sum_{k=1}^{n}(z_{k}+\bar{z}_{k})
    \right)
    \prod_{\ell=1}^{n}\, (z-z_{\ell}) (z - \bar{z}_{\ell})
    \right. \nonumber \\
    &-&
    \left.
    z^{m}
    \,
    \left(
    \bar{z} + \sum_{k=1}^{n}(z_{k}+\bar{z}_{k})
    \right)
    \prod_{\ell=1}^{n}\, (\bar{z}-z_{\ell}) (\bar{z} - \bar{z}_{\ell})
    \right]. \nonumber
\end{eqnarray}
Invoking reasoning we have used in (i), this is further reduced to
\begin{eqnarray}
    \hspace{-1cm}
    \langle q_{2n+1},\, z^{m} \rangle_{S}
    &\propto&
    \int \cdots \int \,
    d^{2}Z_{1} \cdots d^{2}Z_{n} \, d^{2}Z_{n+1} \,
    \prod_{\ell=1}^{n+1}
    \,
    (z_{\ell} - \bar{z}_{\ell})\, w^{2}(z_{\ell},\bar{z}_{\ell})
    \nonumber \\
    &\times&
    \left[
    \det
    \left(
    \matrix{
    1 & z_{1} & \cdots & z_{1}^{2n} & z_{1}^{m} \cr
    1 & \bar{z}_{1} & \cdots & \bar{z}_{1}^{2n}& \bar{z}_{1}^{m} \cr
    \vdots & \vdots & & \vdots & \vdots \cr
    1 & z_{n+1} & \cdots & z_{n+1}^{2n}& z_{n+1}^{m} \cr
    1 & \bar{z}_{n+1} & \cdots & \bar{z}_{n+1}^{2n}& \bar{z}_{n+1}^{m}
    }
    \right)
    \sum_{\ell=1}^{n+1}(z_{\ell} + \bar{z}_{\ell})
    \right. \nonumber \\
    &-&
    \left.
    \det
    \left(
    \matrix{
    1 & z_{1} & \cdots & z_{1}^{2n} & z_{1}^{m+1} \cr
    1 & \bar{z}_{1} & \cdots & \bar{z}_{1}^{2n}& \bar{z}_{1}^{m+1} \cr
    \vdots & \vdots & & \vdots & \vdots \cr
    1 & z_{n+1} & \cdots & z_{n+1}^{2n}& z_{n+1}^{m+1} \cr
    1 & \bar{z}_{n+1} & \cdots & \bar{z}_{n+1}^{2n}& \bar{z}_{n+1}^{m+1}
    }
    \right)
    \right]. \nonumber
\end{eqnarray}
The latter trivially vanishes for $0 \le m \le 2n-1$. This
completes our proof of Eq. (\ref{odd-matrix}).

One may also verify that the normalisation $r_n$ in Eqs.
(\ref{skew-orthogonality-a}) and (\ref{skew-orthogonality-b}) is
related to $A_{n}$ [Eq. (\ref{An})] as $r_n = A_{n+1}/A_{n}$.

\subsection{{\bf Gaussian weight}}
While the representations obtained above are fairly useful to
study, e.g., asymptotic properties of general skew-orthogonal
polynomials and address the issue of universality of eigenvalue
correlations in non-Hermitean random matrix theory at $\beta=4$,
there is no need to resort to them for a simple Gaussian weight
\footnote[5]{This weight may be thought of as originating from the
matrix model $H = H_1 + iv H_2$, with each of $H_\sigma
(\sigma=1,2)$ being drawn from statistically independent Gaussian
symplectic ensembles of Hermitean random matrices $P[H_\sigma]
\propto \exp\left\{ - [N/(1+\tau^2)] \; {\rm Tr}
(H_\sigma^2)\right\}$; the parameter $v^2 = (1-\tau)/(1+\tau)$
[see, e.g., Fyodorov, Khoruzhenko and Sommers (1997)].}
\begin{eqnarray}
    \label{weight}
    w^2_G(z,\bar{z}) = \exp
    \left[
    - \frac{N}{1-\tau^2}
    \left(
    z\bar{z} - \frac{\tau}{2}
    (z^2 + \bar{z}^{2})
    \right)
    \right]
\end{eqnarray}
that we will be interested in in what follows.

In this case, skew-orthogonal Hermite polynomials are simple:
\begin{eqnarray}
    \label{skew-answer-odd}
    q_{2k+1}(z) = \left(
    \frac{\tau}{2N}
    \right)^{k+1/2}
    H_{2k+1} \left(
    z\sqrt{\frac{N}{2\tau}}
    \right), \\
    \label{skew-answer-even}
    q_{2k}(z) = \left(\frac{2}{N}\right)^{k} k!
    \,
    \sum_{\ell=0}^{k} \left(
    \frac{\tau}{2}
    \right)^{\ell} \frac{1}{(2\ell)!!} H_{2\ell}\left(
    z\sqrt{\frac{N}{2\tau}}
    \right).
\end{eqnarray}
Here $H_k(z)$ are `conventional' Hermite polynomials
\begin{eqnarray}
    \label{Hermite}
    H_{k}(z) = \frac{2^{k}}{\sqrt{\pi}}
    \int_{-\infty}^{+\infty}
    dt \, e^{-t^{2}} (z+it)^{k} \nonumber
\end{eqnarray}
orthogonal in the complex plane $z$ with respect to the measure
$w^2(z,\bar{z})\, d^2 Z$ (Di Francesco {\it et al}, 1994):
\begin{eqnarray}
    \label{orth=2}
    \hspace{-1cm}
    \int d^2 Z \, w^2_{G}(z,\bar{z}) \, H_k \left(
    z\sqrt{\frac{N}{2\tau}}
    \right) H_\ell \left(
    \bar{z} \sqrt{\frac{N}{2\tau}}
    \right) = \frac{\pi (1-\tau^2)^{1/2}}{N}
    \frac{2^k k!}{\tau^k} \delta_{k\ell}. \nonumber
\end{eqnarray}
Indeed, straightforward calculation in Eqs.
(\ref{skew-orthogonality-a}) and (\ref{skew-orthogonality-b})
confirms that skew-orthogonality is met with
\begin{eqnarray}
    \label{r}
    r_{k} =2\pi (1-\tau)^{3/2} ({1+\tau})^{1/2}
    \frac{(2k+1)!}{N^{2k+2}}.
\end{eqnarray}

Yet another, integral representation for $q_{2k}(z)$ holds which
is more suitable for our purposes. To derive it we
introduce the function
\begin{eqnarray}
    \label{b1}
    F_{k}(z) = \sum_{\ell=0}^{k} \left( \frac{\tau}{2}\right)^{\ell}
    \frac{1}{(2\ell)!!} H_{2\ell}(z)
\end{eqnarray}
and notice that it satisfies the differential equation
\begin{eqnarray}
    (1+\tau) \frac{\partial F_{k}}{\partial z}
    -
    2 \tau z F_{k}(z)
    =
    - \tau
    \left( \frac{\tau}{2} \right)^{k}
    \frac{1}{(2k)!!} H_{2k+1}(z). \nonumber
\end{eqnarray}
The latter is readily verified by making use of the identity
$H^{\prime}_{\ell}(z) = 2 \ell H_{\ell-1}(z)$. Integrating out, we
infer
\begin{eqnarray}
    \hspace{-2cm}
    F_{k}(z) = \exp\left({\frac{\tau z^{2}}{1+\tau}}\right)
    \bigg[
    \sigma_{k} -
    \frac{\tau}{1+\tau}
    \left(
    \frac{\tau}{2}
    \right)^{k}
    \frac{1}{(2k)!!}
    \int_{0}^{z}dz^{\prime}
    \,
    \exp\left(-\frac{\tau {z^{\prime}}^{2}}{1+\tau}\right)
    H_{2k+1}(z^{\prime})
    \bigg] \nonumber
\end{eqnarray}
where
\begin{eqnarray}
    \sigma_{k} = \sum_{\ell=0}^{k} \left(\frac{\tau}{2}\right)^{\ell}
    \frac{H_{2\ell}(0)}{(2\ell)!!}, \;\;\; H_{2\ell}(0) = (-1)^{\ell}
    \frac{(2\ell)!}{\ell !}. \nonumber
\end{eqnarray}
Summation over $\ell$ can be performed explicitly resulting in
\begin{eqnarray}
    \sigma_{k} = \frac{1}{\sqrt{1+\tau}} \left[
    1 - \frac{\tau^{k+1}}{2^{2k+2} k!} H_{2k+2}(0)
    \int_{0}^{1}\frac{d\xi \, \xi^{k}}{\sqrt{1+\tau \xi}}
    \right]. \nonumber
\end{eqnarray}
Taken together with Eqs. (\ref{skew-answer-even}) and (\ref{b1}),
this brings us to an exact integral representation for the
even-order skew-orthogonal polynomials:
\begin{eqnarray}
    \hspace{-2cm}
    \label{b-exact}
    q_{2k}(z) =  \left( \frac{2}{N}\right)^{k} \frac{k!}{\sqrt{1+\tau}}
\,
    \exp\left({\frac{N z^{2}}{2(1+\tau)}}\right)
    \nonumber \\
    \hspace{-1cm}
    \times
    \bigg\{
    \left[
    1 - \frac{\tau^{k+1}}{2^{2k+2} k!} H_{2k+2}(0)
    \int_{0}^{1}\frac{d\xi \, \xi^{k}}{\sqrt{1+\tau \xi}}
    \right]
    -
    \frac{\tau}{\sqrt{1+\tau}}
    \nonumber \\
    \hspace{-1cm}
    \times
    \left(
    \frac{\tau}{2}
    \right)^{k}
    \frac{1}{(2k)!!}
    \,
    \sqrt{\frac{N}{2\tau}}
    \int_{0}^{z} dw
    \,
    \exp\left(-\frac{N w^{2}}{2(1+\tau)}\right)
    H_{2k+1}\left( w \sqrt{\frac{N}{2\tau}} \right)
    \bigg\}.
\end{eqnarray}

\section{Eigenvalue correlations in $\beta=4$ Gaussian ensembles}
In this section, we apply our findings to explicitly work out
$n$-point correlation function for $\beta=4$ non-Hermitean random
matrix ensemble associated with the Gaussian weight
$w^2_G(z,\bar{z})$ [Eq. (\ref{weight})]. By letting the
parameter $\tau$ tend to zero,
a strongly non-Hermitean Ginibre's ensemble is recovered. Scaling
$\tau$ with matrix size $N$ as $\tau= 1 - \alpha^2/2N$ where
$\alpha \sim O(1)$ one accesses a regime of weak non-Hermiticity
(Fyodorov, Khoruzhenko and Sommers, 1997, 1997a) which is known to
coincide with a zero-dimensional sector of a supersymmetry theory
of disordered systems with a direction (Efetov, 1997).

For other papers addressing non-Hermitean Gaussian ensembles of
symplectic symmetry by field-theoretic ($\sigma$-model) techniques
see, e.g., a supersymmetry treatment by Kolesnikov and Efetov
(1999) and a replica approach by Nishigaki and Kamenev (2002).
Unfotunately, both techniques run into obstacles when one attempts
to study $n$-point correlation function whilst replica
$\sigma$-models seem to reliably provide asymptotic expansions
only (Verbaarschot and Zirnbauer (1985), Kanzieper (2001)).

\subsection{\bf {Finite-$N$ solution}}
In accordance with Eq. (\ref{cf4b=4}), the prekernel [Eq.
(\ref{kernel4b=4})] is the only entity needed to evaluate
$n$-point correlation function. Equations (\ref{prekernel-skew}),
(\ref{skew-answer-odd}), (\ref{skew-answer-even}) and (\ref{r})
furnish the desired solution
\begin{eqnarray}
    \label{finite-N-prekernel}
    \hspace{-2cm}
    \kappa_{N}(z,z^{\prime}) &=& \frac{1}{2\pi} \left(
    \frac{1}{1+\tau}\right)^{1/2}
    \left(
    \frac{N}{1-\tau}
    \right)^{3/2}
    \left[
    \sum_{k=0}^{N-1}\left(
    \frac{\tau}{2}
    \right)^{k+1/2}
    \frac{1}{(2k+1)!!}
    \right.
    \nonumber \\
    \hspace{-2cm}
    &\times&
    \left.
    H_{2k+1} \left(z
    \sqrt{\frac{N}{2\tau}}
    \right)
    \sum_{\ell=0}^{k}
    \left(
    \frac{\tau}{2}
    \right)^{\ell} \frac{1}{(2\ell)!!}
    H_{2\ell}\left(
    z^{\prime} \sqrt{\frac{N}{2\tau}}
    \right)
    - (z  \leftrightarrow z^{\prime})
    \right].
\end{eqnarray}
It holds for arbitrary finite $N$.

\subsection{{\bf Limit of infinite matrices: $N\rightarrow \infty$}}
The large-$N$ limit is different for weakly and strongly
non-Hermitean regimes.

\subsubsection{{\bf Strong non-Hermiticity.}} As $\tau\rightarrow 0$,
the
prekernel simplifies to
\begin{eqnarray}
    \label{prk}
    \hspace{-1cm}
    \kappa_N\left(z,
    z^\prime
    \right)
    =
    \frac{N^{3/2}}{2\pi} \sum_{k=0}^{N-1}
    \left[
    \frac{(z\sqrt{N})^{2k+1}}{(2k+1)!!}
    \sum_{\ell=0}^k \frac{(z^\prime \sqrt{N})^{2\ell}}{(2\ell)!!}
    - (z  \leftrightarrow z^{\prime})
    \right].
\end{eqnarray}
We are interested in a thermodynamic limit $N\rightarrow \infty$
with a blown-up energy resolution $z \mapsto z/\delta_N$ where
$\delta_N = (N/2\pi)^{1/2}$. To this end we have to evaluate
\begin{eqnarray}
    \label{strong4}
    \hspace{-1.3cm}
    \lim_{N\rightarrow \infty}
    \frac{1}{\delta_N^3}
    \kappa_N\left(\frac{z}{\delta_N},
    \frac{z^\prime}{\delta_N}
    \right) =
    \sqrt{2\pi} \nonumber \\
    \times \sum_{k=0}^\infty \sum_{\ell=0}^k
    \left[
    \frac{(z\sqrt{2\pi})^{2k+1}}{(2k+1)!!}
    \frac{(z^\prime\sqrt{2\pi})^{2\ell}}{(2\ell)!!} -
    (z \leftrightarrow z^\prime)
    \right].
\end{eqnarray}
An extra power of $\delta_N$ in the denominator of the l.h.s. is
brought about by a prefactor $(\bar{z}-z)^{1/2}(\bar{z}^\prime -
z^\prime)^{1/2}$ in Eq. (\ref{kernel4b=4}).

Double summation in Eq. (\ref{strong4}) can be performed
explicitly. Denoting
\begin{eqnarray}
    \sigma(z,z^\prime) = \sum_{k=0}^\infty \sum_{\ell=0}^k
    \left[
    \frac{z^{2k+1}}{(2k+1)!!}
    \frac{(z^\prime)^{2\ell}}{(2\ell)!!} - (z \leftrightarrow z^\prime)
    \right], \nonumber
\end{eqnarray}
we observe that
\begin{eqnarray}
    \frac{\partial \sigma}{\partial z} = z\, \sigma +
    e^{zz^\prime}, \;\;\;\;\;
    \frac{\partial \sigma}{\partial z^\prime} = z^\prime\, \sigma
    -
    e^{zz^\prime}. \nonumber
\end{eqnarray}
This suggests that we look for $\sigma(z,z^\prime)$ in the form
\begin{eqnarray}
\sigma(z,z^\prime) = e^{\frac{1}{2}(z^2 + {z^\prime}^2)}
\Lambda(z,z^\prime). \nonumber
\end{eqnarray}
As soon as
\begin{eqnarray}
    \frac{\partial \Lambda}{\partial z} =
e^{-\frac{1}{2}(z-z^\prime)^2}, \;\;\;\;\;
    \frac{\partial \Lambda}{\partial z^\prime} = -
    e^{-\frac{1}{2}(z-z^\prime)^2}, \nonumber
\end{eqnarray}
we obtain
\begin{eqnarray}
\Lambda(z,z^\prime) = \int_0^{z-z^\prime} dt \, e^{-t^2/2} =
\sqrt{\frac{\pi}{2}} \, {\rm erf}\left(
\frac{z-z^\prime}{\sqrt{2}} \right). \nonumber
\end{eqnarray}
This results in
\begin{eqnarray}
    \label{strong=4=integral}
    \hspace{-1cm}
    \lim_{N\rightarrow \infty}
    \frac{1}{\delta_N^3}
    \kappa_N\left(\frac{z}{\delta_N},
    \frac{z^\prime}{\delta_N}
    \right) =
    \pi \exp\left[\pi (z^2 + {z^\prime}^2) \right]
    {\rm erf}\left[
    \sqrt{\pi}(z-z^\prime)
    \right].
\end{eqnarray}
The latter is sufficient to evaluate all $n$-point correlation
functions by means of Eqs. (\ref{cf4b=4}) and (\ref{kernel4b=4}).
For instance, the scaled density of states [Eqs. (\ref{magnify})
and (\ref{r1=4})] reads
\begin{eqnarray}
    \rho_1^{(4)}(z) &=&
    (\bar{z} - z)
    \lim_{N\rightarrow \infty}
    \frac{1}{\delta_{N}^{3}}
    w^{2} \left(
    \frac{z}{\delta_{N}},\frac{\bar{z}}{\delta_{N}}
    \right)
    \kappa_{N}\left(
    \frac{z}{\delta_{N}},\frac{\bar{z}}{\delta_{N}}
    \right) \nonumber \\
    &=&
    8 \pi Y^{2} \exp(-4\pi Y^2)
    \int_0^1\, d\lambda \,\exp(4\pi Y^2 \lambda^2),
\end{eqnarray}
$Y = \Im{\rm m}\, z$. A particular rescaling used in Eq.
(\ref{strong4}) has been chosen in such a way that the scaled
level density $\rho_1^{(4)}(z)$ approaches unity at infinity, $|Y|
\rightarrow \infty$.

\subsubsection{{\bf Weak non-Hermiticity.}}
The regime of weak non-Hermiticity is of a particular interest due
to its close relation to Efetov's model of disordered systems with
a direction. We reiterate that a degree of (weak) non-Hermiticity
is governed by a parameter $\tau$ which scales with the matrix
size $N$ as $\tau = 1 - \alpha^{2}/2N$, $\alpha \sim O(1)$.

The large-$N$ limit of the sum Eq. (\ref{prekernel-skew}) [or Eq.
(\ref{finite-N-prekernel})] is dominated by contributions of terms
with $k$ such that $k/N \sim O(1)$. One therefore needs the
asymptotics of skew orthogonal polynomials $q_k(z)$ at large
indices $k$.

Asymptotics for odd-order skew-orthogonal polynomials $q_{2k+1}$ are
those of
Hermite polynomials $H_{2k+1}$ [Eq. (\ref{skew-answer-odd})].
Utilising the result from standard reference book by Szeg\"o (1939)
\begin{eqnarray}
    H_{2k+1}\left(
    \frac{z}{2\sqrt{k}}
    \right) \simeq \frac{2^{2k+1} (-1)^{k} k!}{\sqrt{\pi k}}\, \sin\, z,
    \nonumber
\end{eqnarray}
we conclude that
\begin{eqnarray}
    \label{odd-ops}
    q_{2k+1}(z) \simeq \frac{2^{2k+1} (-1)^{k} k!}{\pi^{1/2}}
    \left(
    \frac{\tau}{2N}
    \right)^{k+1/2}
    \sin \left(
    z \sqrt{\frac{2kN}{\tau}}
    \right).
\end{eqnarray}
Here $k/N \sim O(1)$ and $z N \sim O(1)$.

Asymptotics for even-order skew-orthogonal polynomials $q_{2k}$
can be read out from Eq. (\ref{b-exact}). Since
\begin{eqnarray}
    \int_{0}^{1}\frac{d\xi \, \xi^{k}}{\sqrt{1+\tau \xi}} \simeq
    \frac{1}{k\sqrt{1+\tau}}, \;\;\; k \gg 1,
    \nonumber
\end{eqnarray}
we derive
\begin{eqnarray}
    \hspace{-2cm}
    q_{2k}(z) \simeq \left( \frac{2}{N} \right)^{k}
    \frac{k!}{\sqrt{1+\tau}}
    \left[
    1
    -
    \frac{\tau}{\sqrt{1+\tau}}
    \left(
    \frac{\tau}{2}
    \right)^{k+1}
    \frac{1}{(2k+2)!!} H_{2k+2}\left( z \sqrt{\frac{N}{2\tau}} \right)
    \right].
\end{eqnarray}
Applying further the asymptotic formula (Szeg\"o, 1939)
\begin{eqnarray}
    H_{2k}\left(
    \frac{z}{2\sqrt{k}}
    \right)
    \simeq
    \frac{2^{2k}(-1)^{k} k!}{\sqrt{\pi k}} \, \cos \, z,
    \nonumber
\end{eqnarray}
we deduce
\begin{eqnarray}
    \label{even-ops}
    q_{2k}(z)\simeq
    \left( \frac{2}{N} \right)^{k}
    \frac{k!}{\sqrt{1+\tau}}
    \left[
    1
    +
    \frac{(-1)^{k} \tau^{k+1}}{\sqrt{1+\tau} \sqrt{\pi k}}
    \cos \left( z \sqrt{\frac{2 k N}{\tau}} \right)
    \right].
\end{eqnarray}

Equations (\ref{odd-ops}) and (\ref{even-ops}) for skew-orthogonal
Hermite polynomials at $k\gg 1$ make it now possible to evaluate
the large-$N$ prekernel. Substituting the two equations into Eq.
(\ref{prekernel-skew}), and replacing the sum over $k$ by an
integral we come up with
\begin{eqnarray}
    \hspace{-1.6cm}
    \label{weak=4=integral}
    \lim_{N\rightarrow \infty}
    \frac{1}{\delta_{N}^{3}}
    \kappa_N \left(
    \frac{z}{\delta_N}, \frac{z^{\prime}}{\delta_N}
    \right) = - \frac{\pi^{3/2}}{4 \alpha^{3}}
    \int_0^1 \frac{d\lambda}{\lambda}
    e^{-\alpha^2 \lambda^2}
    \sin
    \left[\pi
    (z-z^\prime) \lambda
    \right].
\end{eqnarray}
When taking the limit $N \rightarrow \infty$, the scale
$\delta_{N}$ has been set to $\delta_{N} = N\sqrt{2}/\pi$.

In accordance with Eq. (\ref{cf4b=4}), knowledge of the scaled
prekernel Eq. (\ref{weak=4=integral}) is self-sufficient to have
evaluated all $n$-point correlation functions. For instance, the
density of states reads [Eqs. (\ref{magnify}) and (\ref{r1=4})]
\begin{eqnarray}
    \hspace{-1cm}
    \rho_{1}^{(4)}(z) =  \frac{\pi^{3/2}}{2 \alpha^{3}}
    \, Y \,\exp\,(-\pi^{2}Y^{2}/\alpha^{2})
    \int_{0}^{1} \frac{d\lambda}{\lambda} \,
    \exp(-\alpha^{2}\lambda^{2})\,
    {\rm sinh}[2\pi Y \lambda].
\end{eqnarray}

\section{Conclusions}

A problem of eigenvalue correlations in symplectic ensembles of
non-Hermitean random matrices has exactly been solved by the
method of orthogonal polynomials. In close analogy with $\beta=4$
Hermitean matrix ensembles, the $n$-point correlation function is
given by a quaternion determinant [Eq. (\ref{cf4b=4})] of an
$n\times n$ matrix whose entries are quaternions with an image
given by $2\times 2$ matrices in the form of Eq.
(\ref{kernel4b=4}). To evaluate the latter it is convenient (but
not obligatory) to introduce a set of polynomials which are
skew-orthogonal in the complex plane [Eqs.
(\ref{skew-orthogonality-a}) and (\ref{skew-orthogonality-b})].
The skew-orthogonality set by Eq. (\ref{skew-product}) represents
a natural basis in which calculational technology is most
economic.

In Gaussian random matrix ensembles, the eigenvalue correlations
are described by the prekernel Eq. (\ref{finite-N-prekernel})
which further simplifies down to Eqs. (\ref{strong=4=integral})
and (\ref{weak=4=integral}) for strong and weak non-Hermiticity,
respectively. These results apply not too close to the spectrum
edges, which may also be studied within the current framework.

Remarkably, at $\beta=4$, all $n$-point spectral correlation
functions exhibit a peculiar depletion of eigenvalues along the
real axis $\Im {\rm m} z_{\ell}=0$, $1 \le \ell \le n$, where
correlations vanish for both arbitrary matrix size $N$ and a
probability measure $w^2(z,\bar{z})$. As for the remaining
nontrivial functional dependence, we expect it to be universal as
well once a thermodynamic limit is taken. Equations
(\ref{even-matrix}) and (\ref{odd-matrix}) will obviously serve as
a proper starting point to address the universality issue in
either the spectrum bulk or near the complex edges of the
eigenvalue support.

\vspace{0.1cm} {\bf Acknowledgments.} I wish to thank P. J.
Forrester and M. B. Hastings for a number of useful comments and
references.

\smallskip
\section*{References}
\begin{harvard}
\item[] Bruijn de N G 1955 On some multiple integrals involving
        determinants, {\it J. Indian Math. Soc.} {\bf 19} 133
\item[] Dyson F J 1972 Quaternion determinants, {\it Helv. Phys. Acta}
        {\bf 45} 289
\item[] Edelman A 1997 The probability that a random real Gaussian
matrix
        has $k$ real eigenvalues, related distributions, and the
circular law,
        {\it J. Mult. Analysis} {\bf 60} 203
\item[] Efetov K B 1997 Quantum disordered systems with a
        direction, {\it Phys. Rev. B} {\bf 56} 9630
\item[] Eynard B 2001 Asymptotics of skew orthogonal polynomials,
        \JPA {\bf 34} 7591
\item[] Francesco Di F, Gaudin M, Itzykson C and Lesage F 1994
        Laughlin wave-functions, Coulomb gases and expansions of the
        discriminant, {\it Int. J. Mod. Phys.} {\bf A 9} 4257
\item[] Fyodorov Y V, Khoruzhenko B A and Sommers H-J 1997
        Almost Hermitean random matrices: Eigenvalue density in
        the complex plane, \PL {\bf A 226} 46
\item[] Fyodorov Y V, Khoruzhenko B A and Sommers H-J 1997a
        Almost Hermitean random matrices: Crossover from
        Wigner-Dyson to Ginibre eigenvalue statistics,
        \PRL {\bf 79} 557
\item[] Fyodorov Y V, Khoruzhenko B A and Sommers H-J 1998
        Universality in the random matrix spectra in the regime of weak
        non-Hermiticity, {\it Ann. Inst. Henri Poincare (Physique
        Theorique)} {\bf 68} 449
\item[] Ginibre J 1965  Statistical ensembles of complex, quaternion,
and
        real matrices, \JMP {\bf 6} 440
\item[] Halasz M A, Osborn J C and Verbaarschot J J M 1997 Random matrix
        triality at nonzero chemical potential, {\it Phys. Rev. D}
        {\bf 56} 7059
\item[] Hastings M B 2000 Fermionic mapping for eigenvalue correlation
        functions of weakly non-Hermitian symplectic ensemble,
        {\it Nucl. Phys. B} {\bf 572} 535
\item[] Kanzieper E and Freilikher V 1999 Spectra of large random
matrices: A method
        of study, in: {\it Diffuse Waves in Complex Media}, ed. by
        J.-P. Fouque, {\it NATO ASI, Series C (Math. and Phys.
        Sciences)} {\bf 531} 165 (Kluwer, Dordrecht)
\item[] Kanzieper E 2001 Random matrix theory and the replica
        method, {\it Nucl. Phys. B} {\bf 596} 548
\item[] Kolesnikov A V and Efetov K B 1999 Distribution of complex
eigenvalues
        for symplectic ensembles of non-Hermitian matrices, {\it Waves
        in Random Media} {\bf 9} 71
\item[] Lehmann N and Sommers H-J 1991 Eigenvalue statistics of random
real matrices,
        \PRL {\bf 67} 941
\item[] Mahoux G and Mehta M L 1991 A method of integration over
        matrix variables: IV, {\it J. Phys. I France} {\bf 1} 1093
\item[] Mehta M L 1967 {\it Random Matrices} (New York: Academic Press)
\item[] Mehta M L and Gaudin M 1960 On the density of eigenvalues
        of a random matrix, \NP {\bf B 18} 420
\item[] Nishigaki S M and Kamenev A 2002 Replica treatment of
        non-Hermitian disordered Hamiltonians, \JPA {\bf 35} 4571
\item[] Oas G 1997 Universal cubic eigenvalue repulsion for random
        normal matrices, {\it Phys. Rev. E} {\bf 55} 205
\item[] Szeg\"o G 1939 {\it Orthogonal Polynomials}
        (Providence: AMS) \item[] Tracy C A and Widom H 1998 Correlation
        functions,
        cluster functions and spacing distributions for random matrices,
        {\it J. Stat. Phys.} {\bf 92} 809
\item[] Verbaarchot J J M and Zirnbauer M R 1985 Critique of the replica
trick, \JPA {\bf 18} 1093
\smallskip
\end{harvard}
\end{document}